\documentclass[sigconf]{acmart}
\AtBeginDocument{}

\copyrightyear{2026}
\acmYear{2026}
\setcopyright{usgov}
\acmConference[ICS '26]{2026 International Conference on Supercomputing}{July 06--09, 2026}{Belfast, United Kingdom}
\acmBooktitle{2026 International Conference on Supercomputing (ICS '26), July 06--09, 2026, Belfast, United Kingdom}
\acmDOI{10.1145/3797905.3807872}
\acmISBN{979-8-4007-2522-7/2026/07}

\usepackage{subcaption}
\usepackage{algorithm}
\usepackage{algpseudocode}
\usepackage{enumitem}

\begin{document}

\title{StreamGuard: Low-Overhead Resilience for \\ Real-time HPC Data Streams}

\author{Hai Duc Nguyen}
\email{hai.nguyen@anl.gov}
\orcid{0000-0003-4177-0493}
\affiliation{\institution{Argonne National Laboratory}
  \city{Lemont}
  \state{IL}
  \country{USA}
}

\author{Bogdan Nicolae}
\email{bnicolae@anl.gov}
\orcid{0000-0002-0661-7509}
\affiliation{\institution{Argonne National Laboratory}
  \city{Lemont}
  \state{IL}
  \country{USA}
}

\author{Tekin Bicer}
\email{tbicer@anl.gov}
\orcid{0000-0002-8428-5159}
\affiliation{\institution{Argonne National Laboratory}
  \city{Lemont}
  \state{IL}
  \country{USA}
}

\author{Amal Gueroudji}
\email{agueroudji@anl.gov}
\orcid{0009-0004-4830-3139}
\affiliation{\institution{Argonne National Laboratory}
  \city{Lemont}
  \state{IL}
  \country{USA}
}

\author{Matthieu Dorier}
\email{mdorier@anl.gov}
\orcid{0000-0001-9293-2021}
\affiliation{\institution{Argonne National Laboratory}
  \city{Lemont}
  \state{IL}
  \country{USA}
}

\author{Kyle Chard}
\email{chard@uchicago.edu}
\orcid{0000-0002-7370-4805}
\affiliation{\institution{University of Chicago}
  \city{Chicago}
  \state{IL}
  \country{USA}
}

\author{Ian Foster}
\email{foster@uchicago.edu}
\orcid{0000-0003-2129-5269}
\affiliation{\institution{University of Chicago and Argonne National Laboratory}
  \city{Chicago}
  \state{IL}
  \country{USA}
}

\begin{abstract}
  Real-time scientific workflows operate on continuous data streams and must produce timely, high-quality results despite executing on complex, failure-prone infrastructure. Hardware faults, network disruptions, and performance anomalies caused by resource contention or system heterogeneity can severely degrade performance and violate real-time constraints.
  We focus on strengthening the resilience of the producer–consumer streaming pattern, a fundamental building block of scientific streaming workflows. We present two complementary techniques: (i) a dynamic, asynchronous, non-blocking checkpointing mechanism that preserves progress without interrupting computation, and (ii) a progress-aware load redistribution strategy that detects slow workers and proactively rebalances tasks. Together, these mechanisms maintain forward progress and balanced execution even in highly error-prone environments. Experimental results show that our approach reduces the impact of failures and performance anomalies by up to 6$\times$, while introducing less than 1\% overhead in failure-free execution.
\end{abstract}




\maketitle

\thispagestyle{plain}
\pagestyle{plain}

\section{Introduction}
\label{sec:introduction}

Modern scientific instruments
generate large volumes of data as a continuous stream that must be processed in near real-time to guide ongoing experiments. Examples include synchrotron radiation facilities \cite{Thibault:08,dierolf2010ptychographic, liu2020tomogan, Laue_2023}, large-scale sensor deployments \cite{kim2022goal}, and online scientific simulations \cite{yan2025mofa, habib2013hacc}, where data are continuously acquired and processed using multi-stage workflows to produce timely feedback. Unlike traditional batch-oriented scientific workflows, these applications operate on unbounded data streams and must continuously produce meaningful results within strict time constraints \cite{babu2023deep,huerta2019enabling}. As a result, scientific streaming workflows have emerged as a critical paradigm for enabling real-time scientific discovery \cite{nguyen2020motivating, nicolae2024diaspora}.

However, achieving reliable execution for such workflows remains challenging. These workflows are typically deployed on large-scale, distributed infrastructures where failures and performance variability are unavoidable. Fail-stop events, such as process crashes or node failures, interrupt execution and can force costly recomputation if progress is not preserved. Additionally, even if processes do not abruptly end due to fail-stop events, they may be subject to performance degradation
due to various other anomalies: communication delays and timeouts, transient
interference due to competition for shared resources, misconfigured software stacks,
etc. Even if such anomalies are localized, their effect can quickly propagate in the workflow due to dependencies, leading to quality-of-service loss and/or increase in
end-to-end latency. These challenges become increasingly severe as the system scale grows, where failures and performance variability become the norm rather than exception \cite{cappello2014toward, llama3.1meta}.

\paragraph{\bf Limitations of state-of-the-art.}
Traditional HPC resilience techniques emphasize global checkpointing and recovery \cite{dongarra2015fault}, which is appropriate for tightly coupled, bulk-synchronous simulations. In this case, there is an implicit assumption that a safe point can be eventually reached (e.g., end of iteration), where a global barrier ensures there are no more in-flight messages, and simplifies the capture of a globally consistent state as a set of individual checkpoints corresponding to each process. However, stream processing workflows are typically modeled using producer-consumer patterns that are loosely coupled. Since the data are constantly in motion, identifying an opportunity to capture individual checkpoints may not be possible or may introduce excessive coordination overheads. Conversely, many stream-processing frameworks, such as Flink~\cite{carbone2015apache}, are not designed to run on HPC machines. They provide built-in resilience techniques tightly coupled with their execution models~\cite{carbone2015apache,toshniwal2014storm} and are often too conservative (e.g., based on Chandy-Lamport~\cite{chandy1985distributed} algorithm), thereby
incurring high overheads. This limits flexibility and makes them difficult to adopt in specialized scientific workflows. Furthermore, such techniques typically optimize for throughput \cite{noghabi2017samza} or per-partition latency \cite{googlecloud_dataflow} rather than prioritizing end-to-end latency. Thus, there is a need to develop dedicated resilience solutions that overcome these limitations.

\paragraph{\bf Key insights and contributions.} In this paper, we address this gap by rethinking resilience from the perspective of scientific streaming workflows. Instead of enforcing workflow-wide coordination or embedding resilience into a specific framework, we design resilience mechanisms as modular components operating at the producer–consumer level. This design leverages the embarrassingly parallel nature of scientific streaming workloads, allowing failures and performance anomalies to be handled independently with minimal coordination. By decoupling resilience from execution frameworks and avoiding global synchronization, our approach enables resilience to be introduced without sacrificing scalability or real-time performance.

We present StreamGuard, a resilience solution for real-time scientific streaming with two complementary techniques: (i) \textit{dynamic, asynchronous, non-blocking} checkpointing for low-overhead failure recovery, and (ii) \textit{progress-aware load redistribution} that detects slowdown anomalies and rebalances data to satisfy real-time constraints. While each builds on established ideas, our contribution is their integration into a \textit{unified}, \textit{modular} design at the \textit{producer–consumer boundary}, jointly mitigating failures and slowdowns without the workflow-wide coordination required by prior approaches.

We evaluate StreamGuard using a real-world scientific streaming workflow for tomographic reconstruction used at synchrotron radiation facilities \cite{bicer2017trace,bicer2020tomographic}. Experimental results demonstrate that StreamGuard reduces the impact of failures and slowdown anomalies by up to $6\times$ while introducing less than 1\% overhead in failure-free execution. Across a wide range of scales, failure rates, and anomaly conditions, StreamGuard's design consistently maintains processing time close to ideal execution and enables workflows to meet real-time deadlines where existing approaches fail. We summarize our contributions as follows:

\begin{enumerate}[topsep=0pt,itemsep=0pt,leftmargin=*]
\item
A unified, decoupled producer–consumer–centric resilience design that integrates dynamic non-blocking checkpoint-and-retry with progress-aware load balancing as composable modules, avoiding workflow-wide coordination, reducing performance degradation from failures and slowdowns by up to $4\times$ and $5\times$, respectively (\S\ref{sec:solution}).
\item A modular resilience architecture for real-time scientific streaming that allows applications to selectively enable resilience mechanisms as needed. When combined, these modules effectively prevent compounded performance degradation, with performance increase to at most 40\% under extreme failure and anomaly conditions at scale (\S\ref{sec:implementation}).
\item A systematic evaluation using a real-world scientific streaming application that demonstrates improved robustness, scalability, and performance transparency compared to state-of-the-art HPC and stream-processing resilience approaches,
while consistently meeting real-time constraints in scenarios where existing solutions may fail (\S\ref{sec:evaluation}).
\end{enumerate}

\section{Background}
\label{sec:background}

\begin{figure}
    \centering
    \includegraphics[width=\linewidth, trim={40 189 55 240}, clip]{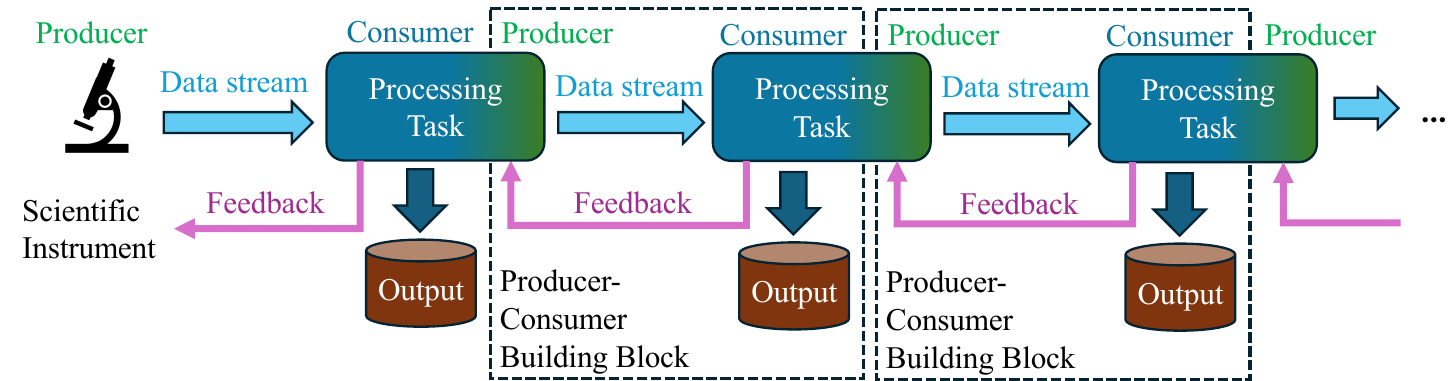}
    \caption{Scientific streaming workflow is formed by a sequence of producer-consumer building blocks}
    \label{fig:scientific-streaming-producer-consumer}
    \vspace{-0.1in}
\end{figure}

We consider a large class of streaming scientific applications that follow a producer-consumer pattern as illustrated in \autoref{fig:scientific-streaming-producer-consumer}. The workflow begins at a scientific instrument, such as a synchrotron radiation facility, that continuously generates large volumes of data during experiments. These data are transferred to HPC systems for processing through a sequence of stages, where each stage consists of multiple parallel workers processing different portions of the data stream (i.e., partitions).

\subsection{Producer-Consumer Streaming}
\label{sec:background-producer-consumer}

The communication between the stages follows a producer–consumer model. Upstream workers act as \textit{producers} by emitting data records, while downstream workers act as \textit{consumers} that receive and further process incoming data records. Depending on application requirements, the data records may be distributed among consumers through different patterns, such as shuffle, where records are scattered across consumers, or broadcast, where every record is delivered to all consumers. Regardless of the distribution pattern, data flowing between producers and consumers is typically divided into sequential \textit{partitions} with unique identifiers (or keys). These partitions can be processed independently, enabling efficient parallel execution over large-scale data streams.

After processing, consumers may generate intermediate outputs, persist state, or provide feedback to upstream stages. The outputs are then forwarded downstream, where the same producer–consumer interaction repeats. Because most computation and data movement occur at these boundaries, the producer–consumer pair forms the primary unit where performance variability, failures, and anomalies directly affect end-to-end workflow behavior.

\subsection{Key Characteristics}
\label{sec:background-key-characteristics}

Building on the producer–consumer pattern described earlier, scientific streaming workflows exhibit characteristics that distinguish them from both traditional HPC
and stream-processing workloads.

\paragraph{\bf Unbounded data stream.}
Data are continuously generated by instruments or edge devices as an unbounded sequence of records. Each record captures only a small portion of the evolving experiment, and the full outcome emerges gradually as more data arrive. This differs fundamentally from traditional HPC workloads, where the complete dataset is typically available in a parallel file system before processing begins. In scientific streaming, workers must operate on incomplete and continuously evolving data.

\paragraph{\bf Iterative refinement.}
Because data arrive incrementally, early outputs are necessarily approximate. As additional data are consumed, workers iteratively refine their results, steadily improving accuracy and quality. This pattern is common in applications such as iterative reconstruction \cite{nguyen2025resilient, bicer2017trace} and long-term campaigns \cite{barrand2001gaudi,steiner2019organic,li2024unifaas}. Unlike conventional stream-processing systems that handle many independent, small, and fine-grained records, scientific streams often consist of coarse-grained, high-volume data units that collectively represent meaningful scientific states.

\paragraph{\bf Embarrassingly parallel processing.}
Streamed partitions can be processed independently by many parallel workers with minimal coordination. As a result, scientific streaming workflows
often rely on workers deployed across multiple compute nodes, leveraging their collective power to accelerate stream processing.

\paragraph{\bf Real-time quality constraints.}
Although the stream is continuous, processing must produce useful results within practical time limits imposed by system policies or experimental needs. Therefore, the goal is not merely to process all incoming data, but to maximize output quality within a bounded execution window—a notion often captured as Quality of Service (QoS).

\section{Problem Formulation}

\begin{table}[ht]
    \caption{Notations used in this paper}
    \label{tab:notations}

\vspace{-1ex}

    \centering
    \begin{tabular}{c|l}
        \textbf{Notation} & \textbf{Definition} \\ \hline
        $s_{i}$ & A partition of the data stream\\
        $N$ & Number of data partitions\\
        $k_{i}$ & Data partition identifier (or key)\\
        $w_{j}$ & Workers to process data partitions\\
        $M$ & Number of workers\\
        $p_{i}(t)$ & Progress of a data partition $s_{i}$ at a time $t$\\
        $q_{j}(t)$ & Progress of a worker $w_{j}$ at a time $t$\\
        $L$ & Real-time deadline\\
        $P$ & Target progress before reaching the deadline\\
        $\mu$ & Per-worker mean time to failure
    \end{tabular}
\end{table}

This paper addresses the \textbf{resilience challenges} of the producer-consumer pattern. We model data stream $S$ between a producer and a consumer as a set of independent partitions $S = \{s_{1},s_{2},...,s_{N}\}$ where a partition $s_{i}$ is the set of records that need to be processed together in a sequential order (e.g., data collected by a single device over time) identified by a unique key $k_{i}$. Let $w_{1}, \dots, w_{M}$ denote the available \textit{workers} (e.g., processes or compute nodes) allocated to the consumer to process data. We assume $M \leq N$, meaning that each worker may process multiple partitions. We also assume that data partitions have similar computational demand, which can be achieved through appropriate data partitioning methods~\cite{liu2024adaptive, sun2024adaptive}. In contrast, workers are not assumed to be identical, as their performance may vary due to hardware differences, resource contention, or dynamic runtime conditions. To flexibly work around this heterogeneity, we allow data partitions to be reassigned from one worker to another and let $a(s_{i}, w_{j}, t)$ be an indicator function specifying whether the data partition $s_{i}$ is assigned to the worker $w_{j}$ at time $t$:

\begin{equation}
    a(s_{i}, w_{j}, t) = \begin{cases}
        1, & \text{if } s_{i} \text{ is assigned to } w_{j} \text{ at } t\\
        0, & \text{otherwise}
    \end{cases}
\end{equation}
We assume time is discretized into time units in which partition-to-worker assignment remain unchanged.
Since consumer workers process data in an iterative manner, we define \textit{partition progress} $p_{i}(t)$ as the number of records that have been processed over the data partition $s_{i}$ by the time $t$. Likewise, \textit{worker progress}, $q_{j}(t)$, is the total progress a worker $w_{j}$ has made on its assigned partitions:
\begin{equation}
    q_{j}(t) = \sum_{i=1}^{N}\sum_{\tau=1}^{t}(p_{i}(\tau)-p_{i}(\tau-1))a(s_{i}, w_{j},\tau)
    \label{equ:worker-progress}
\end{equation}

Compared to traditional batch processing, streaming workflows can produce usable outputs without waiting for the complete dataset. Initial results can be generated after a few processing iterations and progressively refined as additional data arrive. This property makes streaming well-suited for real-time applications, where consumers must produce meaningful outputs within a specified time window.
We model this real-time requirement using a deadline $L$ and a progress target $P$. The real-time constraint is satisfied if all data partitions reach at least progress level $P$ by the deadline $L$:
\begin{equation}
    \forall i : p_{i}(L) \geq P
    \label{equ:real-time-constraint}
\end{equation}
The notations used in this formulation are summarized in \autoref{tab:notations} and are used throughout the remainder of the paper.

The producer-consumer pattern
often requires massive parallelism to process large data streams, so the workflow can be highly sensitive to failures and performance anomalies. We consider two common classes of disruptions.

\paragraph{\bf Fail-stop events}

A fail-stop event occurs when a worker abruptly stops execution due to errors such as segmentation faults or out-of-memory conditions \cite{fail-stop}. Fail-stop failures are often modeled as random processes \citep{ckpt-overview}, where workers fail independently with identical probabilities. These failures are characterized by the mean time to failure (MTTF), denoted as $\mu$, which represents the average time a single worker runs before failing. Worker failure probability is typically described using an exponential distribution Exp($\lambda$), where $\lambda = \frac{1}{\mu}$ represents the failure rate \citep{dongarra2015fault}. The mean time between failures (MTBF) for the whole consumer is given by:
\begin{equation}
    \mu_{p} = \frac{\mu}{M}\,.
    \label{equ:app-mtbf}
\end{equation}

\autoref{equ:app-mtbf} shows that increasing parallelism reduces MTBF. This is intuitive, as more workers introduce more potential points of failure, causing failures to occur more frequently. This phenomenon poses a significant challenge for achieving resilient execution at large scales. For instance, if the MTTF of a single worker is one day, a consumer with 1000 workers would, on average, fail in less than 2 min---often not long enough for the consumer to provide any meaningful results \cite{bicer2017trace}. Moreover, even with workers running on infrastructures designed to operate reliably for years, the sheer scale of large systems amplifies failure rates. Studies have shown that large-scale deployments, with thousands of compute nodes, can experience failures several times per day due to the aggregated likelihood of individual node failures \cite{cappello2014toward, ferreira2011evaluating}.

\paragraph{\bf Anomalous performance degradation.}
Not all performance incidents result in complete worker failures. In many cases, long-lasting conditions such as network congestion, resource contention, hardware interference, or inherent system heterogeneity reduce worker performance but do not stop execution. As a result, data streams assigned to affected workers progress more slowly than others. We refer to this phenomenon as a \textit{slowdown anomaly}. Because real-time execution requires \textit{all} streams to reach a minimum progress threshold (as defined in \autoref{equ:real-time-constraint}), overall consumer performance becomes dominated by the slowest stream.

Slowdown anomalies can have severe consequences if left unmitigated. Empirical studies of large-scale distributed and cloud systems have shown that worker execution times commonly exhibit heavy-tailed or power-law-like distributions under heterogeneous environments~\cite{dean2013tail,reiss2012heterogeneity}. In such settings, the majority of workers progress at similar speeds, while a small fraction experiences disproportionately longer execution times. This behavior is widely known as the \textit{straggler phenomenon}~\cite{dean2008mapreduce,ananthanarayanan2010reining}, where a few slow workers dominate overall completion time despite representing only a small portion of the system.

The impact of slowdown anomalies is particularly pronounced in scientific streaming workflows. Since outputs are only meaningful when progress across all partitions reaches a required level, the slowest stream effectively determines whether the real-time deadline can be satisfied. As system scale increases, the probability of encountering slow workers also increases, amplifying the likelihood that slowdown anomalies degrade overall performance. These characteristics make anomaly mitigation a critical requirement for maintaining reliable real-time execution in scientific streaming environments.

\section{Related Work}
\label{sec:related-work}

\paragraph{\bf Coordinated Checkpoint-Restart}
HPC resilience is dominated by the checkpoint-restart (C/R) paradigm~\cite{CRSurvey24}, necessitated by the tightly coupled nature of parallel applications. Because such applications typically use MPI for inter-process communication, the failure of a single process invalidates the global state and requires all nodes to roll back to a coordinated recovery point. Recent research has shifted from global blocking checkpoints toward multilevel hierarchies that adapt to failure severity, spanning local memory tiers (GPU HBM, host memory, SSDs), erasure coding, buddy replication, burst buffers, and external object stores such as DAOS~\cite{manubens2024exploring}.
Prominent examples in this space are VELOC~\cite{VELOC-SuperCheck21,VeloCIPDPS19}, SCR~\cite{SCR14}, FTI~\cite{FTI11}, and ADIOS~\cite{adios2_2020}. A key optimization is to enable asynchronous capture and flushing of checkpoints in order to hide I/O overheads and various other transformations (data aggregation~\cite{VELOC-FGCS24}, compression~\cite{LineageComp-HIPC23}, etc.). However, despite extensive work in reducing checkpoint overhead, these approaches implicitly assume a safe point at which a global barrier ensures no in-flight messages exist (e.g., end of iteration). In stream-processing workflows, such a safe point does not naturally exist and forcing one requires expensive synchronization (\S\ref{sec:background-key-characteristics}), making global C/R infeasible.

\paragraph{\bf Resilient Stream Processing}
Conversely, stream-computing resilience leverages the loosely coupled nature of producer–consumer patterns through checkpoint-replay mechanisms derived from the Chandy-Lamport algorithm~\cite{chandy1985distributed}. Unlike HPC's rigid synchronization, stream processors such as Google Cloud Dataflow~\cite{googlecloud_dataflow} and Apache Flink~\cite{carbone2015apache, flink_checkpoint_docs}  inject barrier markers into the data stream to capture asynchronous distributed snapshots that establish a consistent recovery boundary accounting for in-flight messages, and replay input logs on failure for exactly-once semantics. Alternatively, Spark uses lineage-based micro-batch checkpointing \cite{armbrust2018structured,zaharia2013discretized}, while Storm and Heron use tuple acknowledgment and replay \cite{toshniwal2014storm, kulkarni2015twitter}. Data-intensive middleware can cache compact reduction objects and redistribute unfinished work after failures~\cite{bicer2010supporting}. These mechanisms decouple operator recovery from overall latency, supporting high throughput under transient faults. However, such approaches are often too conservative in that checkpoint capture is subject to slow serialization and synchronous I/O, while replay is done in a rigid fashion, without load-balancing or straggler mitigation strategies, which ultimately propagates delays through the entire workflow both during failure-free and faulty regimes of operation.

\paragraph{\bf Mitigation of Non-Fail-Stop Anomalies and Gray Failures}
A critical frontier in resilience research involves the management of non-fail-stop anomalies, often referred to as ``gray failures'', where nodes remain operational but exhibit sub-optimal performance~\cite{huang2017gray}. These issues stem from resource contention, network jitter, or software misconfigurations, and are particularly disruptive because they propagate delays through downstream dependencies, leading to significant tail latencies~\cite{dean2013tail}. Unlike fail-stop errors, these anomalies cannot be resolved by simple restarts. Instead, modern resilience frameworks employ proactive mitigation techniques, such as speculative execution~\cite{dean2008mapreduce,ananthanarayanan2010reining} (launching redundant ``straggler'' tasks) and dynamic load balancing~\cite{lu2021autoflow, song2023sponge, goodrich2023esnet}. By utilizing observability-driven feedback loops, these systems detect performance deviations and rebalance workloads in real-time, preventing a localized slowdown from cascading into a systemic bottleneck. However, such techniques are often oblivious to end-to-end QoS objectives and latency considerations.

\section{StreamGuard: Resilient HPC Streaming}
\label{sec:solution}

We propose \textbf{StreamGuard}, a resilience solution that combines two complementary mechanisms to protect against failures and performance anomalies while preserving the performance characteristics required by real-time scientific streaming workflows.

\subsection{Overall Design}

\begin{figure}
    \centering
    \includegraphics[width=\linewidth, trim={50 190 70 190}, clip]{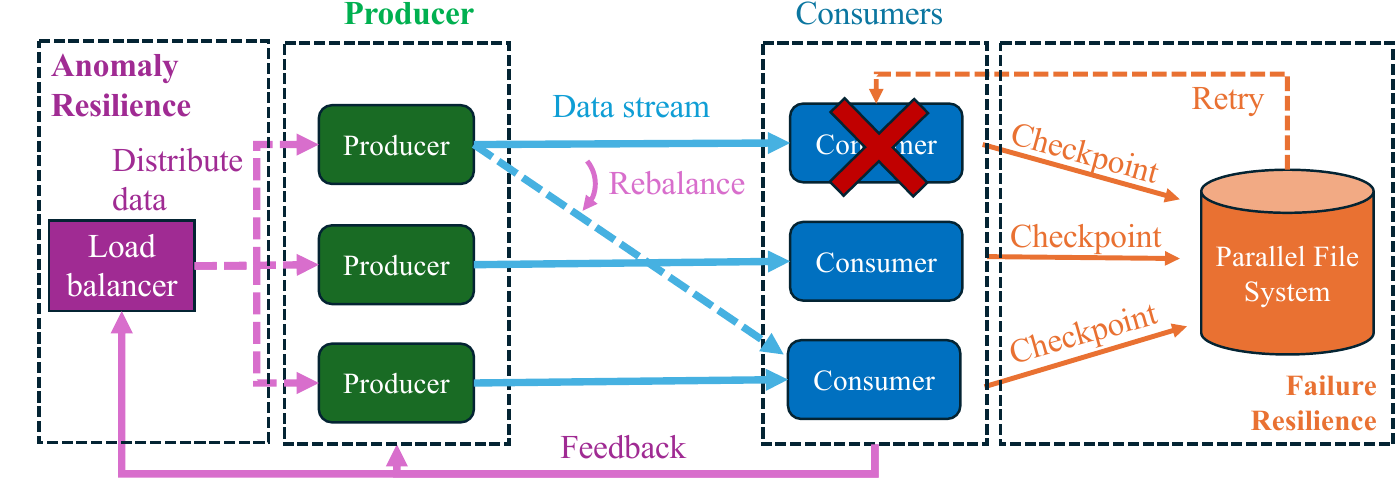}
    \caption{Resilience solutions: (i) Checkpoint and retry for failure resilience and (ii) dynamic load-balancing for anomaly resilience}
    \label{fig:architecture}
\end{figure}

We design resilience mechanisms that decouple resilience from workflow-specific execution frameworks while explicitly targeting real-time scientific streaming requirements.
\autoref{fig:architecture} illustrates the overall architecture of StreamGuard.
The central design principle is \textit{modularity}. Rather than embedding resilience into a specific execution framework or enforcing workflow-wide coordination, resilience is provided as independent modules that can be selectively applied to different parts of a streaming workflow. This design improves portability and allows scientific workflows to integrate resilience support without modifying their execution model.

Instead of operating at the workflow level, we focus on producer–consumer pairs, which form the fundamental building blocks of scientific streaming applications. Most computation and performance sensitivity occur at these boundaries, where data are produced, processed, and forwarded. Operating at this granularity avoids unnecessary global coordination and allows resilience mechanisms to be applied only where needed, reducing overhead while maintaining flexibility.
For each producer–consumer pair, we provide two complementary resilience mechanisms that independently address failures and performance anomalies.

The \textit{failure resilience} mechanism mitigates worker failures through a lightweight checkpoint and retry process. Consumer workers periodically capture snapshots of their processing state and persist them as checkpoints on the parallel file system. Upon failure, the execution module automatically restarts affected workers on available resources and restores the most recent checkpointed state, allowing execution to resume from a consistent point rather than restarting from the beginning. This significantly reduces recovery time and limits disruption to ongoing stream processing.

The \textit{anomaly resilience} mechanism addresses slowdown anomalies caused by performance heterogeneity or resource contention. Using runtime feedback from consumers, the module continuously monitors progress on data partitions and estimates effective worker processing capacity. When imbalance is detected, a load balancer dynamically adjusts partition-to-worker assignments by migrating slower partitions to faster workers. This maintains balanced progress across partitions and helps ensure that end-to-end processing remains within real-time constraints.

Together, these modules provide resilience against both failures and performance anomalies while avoiding tight coupling between mechanisms or reliance on global coordination. This separation prevents failures and slowdowns from compounding and enables robust execution across a wide range of operating conditions. The detailed design and implementation of each module are described in the following subsections.

\subsection{Key Design Principles}

\subsubsection{Decoupled Lightweight Checkpoint-Restart}
\label{sec:solution-ckpt}

We design a lightweight checkpoint-restart mechanism that leverages the embarrassingly parallel structure of scientific streaming workflows.
The design is based on three key principles.

\medskip
\paragraph{\bf Asynchronous, Per-Partition Checkpointing.}
Instead of maintaining a single global checkpoint, we construct checkpoints independently for each data partition $s_{i}$. Each such checkpoint is self-contained and includes all information required to restore processing progress, such as intermediate state and consumed records. Because streams are processed independently, workers can checkpoint without coordinating with other workers. This eliminates global synchronization overhead and simplifies recovery, allowing failed workers to resume execution from their most recent consistent state while unaffected workers continue processing normally.

\begin{figure}
    \centering
    \includegraphics[width=\linewidth, trim={100 185 140 160}, clip]{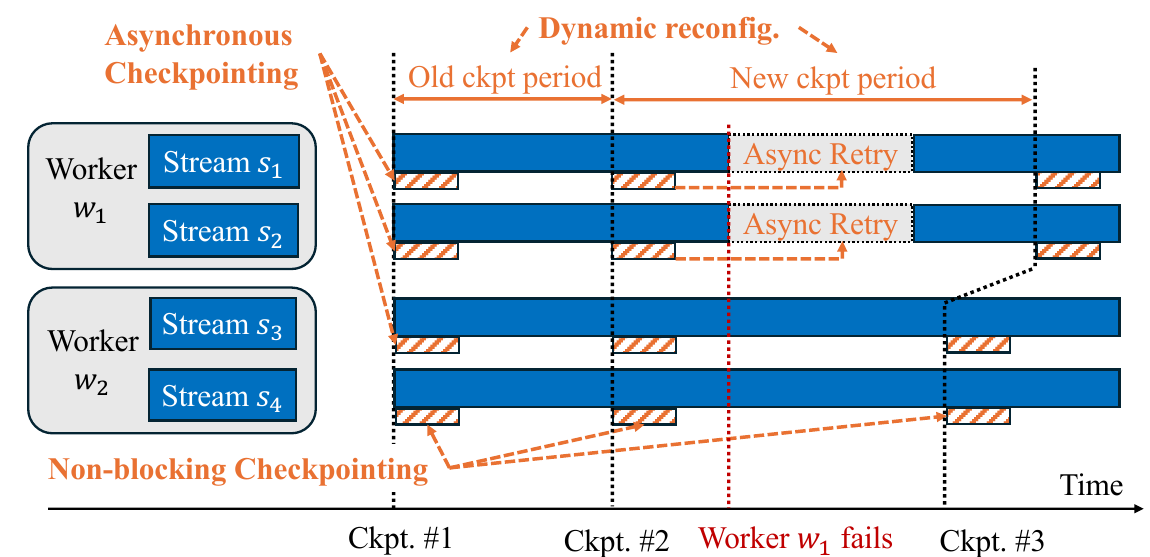}
    \caption{Checkpoint and retry example with two workers and four partitions}
    \label{fig:resilient-solution-ckpt-retry}
\end{figure}

\medskip
\paragraph{\bf Non-blocking Execution.}
Scientific stream processing is typically computation-intensive, whereas checkpoint creation is I/O-intensive. This allows us to remove checkpoint creation from the execution critical path by overlapping the two processes. When a checkpoint is triggered, the worker provides references to the memory regions containing checkpoint data and immediately resumes computation, while a separate process persists the checkpoint in the background. By overlapping computation with I/O, the system improves resource utilization and keeps checkpoint overhead negligible during failure-free execution.

\medskip
\paragraph{\bf Dynamic Checkpoint Period.}
A central challenge in checkpoint-based resilience is determining how frequently checkpoints should be created. Frequent checkpointing reduces recomputation after failures but increases runtime overhead, whereas infrequent checkpointing minimizes overhead but increases recovery cost. Classical solutions such as the Young/Daly formula \cite{daly2006higher, young1974first} provide an optimal checkpoint interval under static assumptions. However, scientific streaming environments often experience fluctuating failure rates and variable I/O performance due to slowdown anomalies.

To address this, we dynamically adjust the checkpoint period for each worker based on observed runtime behavior. Specifically, each worker $w_j$ computes its checkpoint period based on a modified Young/Daly formula:
\begin{equation}
    W(w_{j}) = \sqrt{2 \frac{\mu}{M} C(w_{j})}\,,
    \label{equ:dynamic-ckpt-reconfig}
\end{equation}
where $M$ is the number of workers, $\mu$ the MTTF, and $C(w_{j})$ the empirically observed per-worker checkpoint overhead. Each worker computes $W(w_j)$ independently from its own $C(w_j)$ and reconfigures it after failures or when checkpoint overhead changes, allowing the system to adapt to changing I/O and execution conditions while minimizing checkpoint overhead and recomputation cost.

\medskip
\noindent
\autoref{fig:resilient-solution-ckpt-retry} illustrates the mechanism with four partitions across two workers, each creating its own checkpoint asynchronously in the background. When a worker fails, only the partitions assigned to that worker stop execution, while surviving workers continue processing unaffected. The failed partitions are later recovered using their most recent checkpoints, avoiding costly recomputation. Because checkpoints are independent across streams, recovery requires no inter-worker synchronization, allowing failure handling to remain lightweight and scalable.

\begin{algorithm}[H]
    \caption{Load-balancing Algorithm}
    \label{alg:load-balancing}
    \begin{algorithmic}[1]
        \State $\beta \gets \frac{N}{M}$
        \While{\textbf{true}}
            \State{Load balancing step $t \gets t + 1$}
            \State \text{Update $p_{i}(t)$ for all partitions $s_{i}$}
            \State \text{Calculate $c_{j}$ for all workers $w_{j}$ with \autoref{equ:worker-cap}}
            \State \text{Compute $G(t)$ by \autoref{equ:load-balance-goal}}
            \If{$G(t) < G_{threshold}$}
                \State $s_{min} \gets \arg\min_{s_{i}}{p_{i}}(t)$
                \Comment{Slowest partition}
                \State {$c_{min} \gets$ $s_{min}$'s worker capacity}
                \For{\textbf{all} workers $w_{j}$}
                    \State {$\text{surplus}_{j} \gets c_{j}(t) - \sum_{i}a(s_{i}, w_{j}, t)$}
                \EndFor
                \State {sort $w_{j}$ by decreasing order of $c_{j}$ }
                \State {partition\_moved $\gets$ \textbf{false}}
                \For{each worker $w_j$ in sorted order}
                    \If{surplus$_{j}$ > 1 \textbf{and} $c_{j} > c_{min}$  }
                        \State {Reassign $s_{min}$ to $w_{j}$}
                        \State {partition\_moved $\gets$ \textbf{true}}
                        \State \textbf{break}
                    \EndIf
                \EndFor
                \If{\textbf{not} partition\_moved}
                    \State {$s_{max} = \arg\max_{s_{i}}p_{i}(t)$}
                    \State {swap $s_{min}$ and $s_{max}$}
                \EndIf
            \EndIf
            \State $G_{threshold} \gets G$
        \EndWhile
    \end{algorithmic}
\end{algorithm}

\subsubsection{Load Balancing}
\label{sec:solution-load-balancing}

To mitigate degradation anomalies, we propose a dynamic load-balancing algorithm that continuously tracks per-partition processing progress and redistributes partitions across workers according to their effective processing speed. The objective is to prevent slow workers from violating the real-time deadline.
Consider a producer–consumer pair operating within a real-time deadline window $[0, L]$, where $0$ is the time when the first record is emitted by the producer and $L$ is the processing deadline. Let $p_i(t)$ denote the progress of partition $s_i$ at time $t \in [0, L]$. We define
\[
p_{\min}(t) = \min_i p_i(t), \quad
p_{\max}(t) = \max_i p_i(t)
\]
as the slowest and fastest progress on data partitions at time $t$, respectively. From the real-time constraint (\autoref{equ:real-time-constraint}), the workflow meets the deadline only if the slowest partition completes on time, i.e., $p_{\min}(L) \geq P$.
This implies that overall completion time is determined solely by the slowest partition. Consequently, improving real-time performance requires prioritizing the progress of lagging partitions, even if doing so slightly slows faster ones because the number of available workers is limited to $M$.
To quantify imbalance among partitions, we define the \textit{progress ratio}, $G(t)$, as:
\begin{equation}
    G(t) = \frac{p_{\min}(t)}{p_{\max}(t)},
    \label{equ:load-balance-goal}
\end{equation}
which measures how evenly progress is distributed across partitions. When $G(t)=1$, all partitions advance at the same rate. The goal of load balancing is therefore to maintain $G(t)$ close to 1 by reducing progress disparity among partitions.

We achieve this objective by identifying worker performance dynamically using observed execution behavior. Let the \textit{worker active time} $v_j(t)$ denote the total time during which worker $w_j$ is alive and assigned at least one partition. Let $q_j(t)$ denote the total progress made by that worker (see \autoref{equ:worker-progress}). We define the effective \textit{worker speed} as $r_j(t) = \frac{q_j(t)}{v_j(t)}$, which captures the average processing rate of the worker during execution. Partitions associated with slower workers are candidates for reassignment to faster workers.

However, assigning too many partitions to a fast worker may overload it and degrade overall performance. To avoid this, we introduce a capacity model. Recall that $N$ is the total number of partitions and $M$ the number of workers. The ideal number of partitions per worker under homogeneous conditions is $\beta = \frac{N}{M}$.
We adjust this quantity for worker speed to obtain the worker's capacity, $c_j(t)$:
\begin{equation}
    c_j(t) = \beta
    \frac{r_j(t)}{\frac{1}{M}\sum_{j} r_j(t)},
    \label{equ:worker-cap}
\end{equation}
which represents the number of partitions a worker $w_j$ can sustain relative to the average worker. A slow partition is reassigned only to workers whose current load is below their capacity. If no such worker exists, the algorithm swaps the slowest partition with the fastest one, sacrificing some progress of the fastest stream in order to improve the global progress balance.

The complete procedure is summarized in \autoref{alg:load-balancing}. The algorithm operates periodically: partition progress is updated at fixed intervals, after which the progress ratio $G(t)$ is evaluated. If $G(t)$ falls below a predefined threshold $G_{\text{threshold}}$, reassignment is triggered (Line~\#8). The algorithm selects the slowest partition ($s_{\min}$, Line~\#9) and attempts to migrate it to the fastest worker with available capacity (Lines~\#10–\#21). When no worker has spare capacity, the slowest and fastest partitions are swapped to improve balance. Through repeated adjustments, the algorithm gradually equalizes progress across streams, preventing slowdown anomalies from propagating to end-to-end processing time.

\begin{figure}
    \centering
    \includegraphics[width=\linewidth, trim={40 220 50 220}, clip]{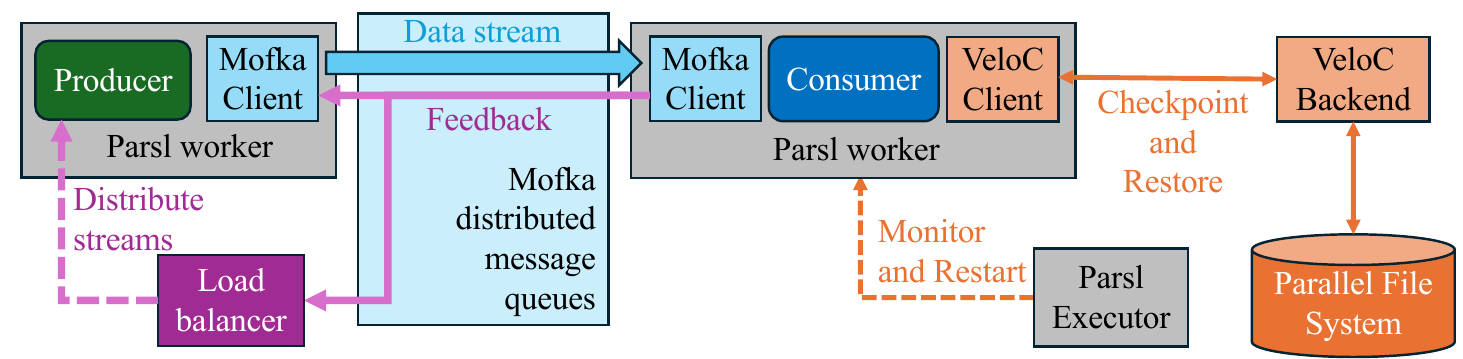}
    \caption{StreamGuard implementation for an individual producer-consumer building block with four resilience modules: (i) resilience execution (gray), (ii) resilience communication (blue), (iii) failure protection (orange), and (iv) slowdown protection (purple)}
    \label{fig:implementation}
\end{figure}

\section{StreamGuard Implementation}
\label{sec:implementation}

We implement StreamGuard as a software development kit (SDK) that provides resilience capabilities as independent, composable modules. Applications enable only the mechanisms they need, while the SDK design avoids tight coupling with specific execution frameworks, allowing integration into existing workflows without modifying their execution model.
The overall architecture of StreamGuard is illustrated in \autoref{fig:implementation} with four resilience modules, each responsible for a distinct aspect of reliable stream processing.

\paragraph{\bf Resilience Execution Module.}
This module provides reliable execution by automatically restarting failed workers on available compute resources. We implement this functionality using Parsl~\cite{babuji2019parsl} (gray boxes in \autoref{fig:implementation}). Each stream processing task is implemented as a function whose inputs are records from a given partition $s_{i}$ and outputs are data to be forwarded downstream. The function is executed inside a Parsl-managed container (i.e., the worker $w_{j}$ in our resilience model), allowing Parsl to monitor worker liveness and automatically relaunch failed workers while preserving identifiers and stream mappings.
The execution module exposes runtime information, including worker status, failure events, and resource availability, to other resilience modules through a resilient communication layer (described below). This information enables checkpointing and load-balancing decisions to adapt dynamically during execution.

\paragraph{\bf Resilience Communication Module.}
This module provides reliable producer–consumer communication via a publish/subscribe distributed queue (blue boxes in \autoref{fig:implementation}), offering two benefits. First, the queue provides reliable transmission with QoS guarantees (e.g., exactly-once), letting streams tolerate network failures without application-level recovery logic. Second, data persistence in the queue is decoupled from worker execution (workers retrieve data only when ready), preventing failures and slowdowns from propagating across the workflow.

We use Mofka \cite{mofka, mofka-tekapp}, a high-performance distributed message queue designed for HPC environments, as the underlying communication layer. StreamGuard provides a wrapper that allows applications to define logical data streams using unique keys, where each key $k_i$ corresponds to a data partition $s_i$ mapped to a dedicated Mofka topic. Producers publish records to these topics, which are subscribed by consumers for transmission. This abstraction simplifies application logic and enables other resilience modules to observe partition progress and exchange runtime information transparently and reliably.

\paragraph{\bf Failure Protection Module.}
This module protects workflow state against failures through the lightweight checkpointing mechanism described in \autoref{sec:solution-ckpt}. The implementation is based on a customized version of VeloC \cite{VELOC-SuperCheck21}, a multi-level checkpoint runtime specialized for HPC infrastructure and large-scale deployment. Checkpointing is organized per producer–consumer pair. Each checkpoint set represents the state of a consumer, where entries correspond to individual data partitions identified by their keys. During initialization, applications register checkpointable state and receive checkpoint handlers that can be invoked during execution.

During normal execution, each worker $w_{j}$ issues checkpoint requests after reaching consistent processing points, typically at the end of an iteration. If the elapsed time since the previous checkpoint exceeds the current checkpoint period $W(w_{j})$, the in-memory state is passed to a background VeloC process and written asynchronously, allowing computation to continue without interruption. Since checkpoints for different partitions are handled independently, the mechanism avoids global synchronization and introduces no inter-worker communication overhead. The module continuously monitors checkpoint latency and observed failure frequency to dynamically adjust the checkpoint period according to Equation~\ref{equ:dynamic-ckpt-reconfig}, maintaining near-optimal overhead.

A worker may also explicitly enforce checkpointing when required by execution semantics, for example, after reaching a critical processing milestone or prior to reassigning a partition to another worker during load balancing. In such cases, the SDK issues a blocking checkpoint request that forces VeloC to complete the checkpoint before control is returned to the worker. This ensures that the worker state is fully persisted before execution proceeds, preventing progress loss during reassignment or recovery.

\paragraph{\bf Slowdown Protection Module.}
This module mitigates slowdown anomalies using the dynamic load-balancing mechanism described in \autoref{sec:solution-load-balancing} (purple boxes in \autoref{fig:implementation}). Its objective is to maintain balanced progress across streams despite heterogeneous worker performance. The module monitors partition progress through the communication module by tracking per-partition record consumption rates at consumer workers and estimating effective processing speed.

Applications may optionally enable this module for selected producer–consumer pairs. When an imbalance is detected, the load balancer initiates partition reassignment by instructing the current consumer to checkpoint and release the affected partition. The partition is then reassigned to a faster worker, which loads the corresponding checkpoint, subscribes to the partition's topic, and resumes processing. Because reassignment involves only the previous and new consumer, migration proceeds without global synchronization or interruption of other streams, minimizing impact on overall workflow performance.

\section{Evaluation}
\label{sec:evaluation}

\begin{figure}
    \centering
    \includegraphics[width=\linewidth, trim={135 220 100 190}, clip]{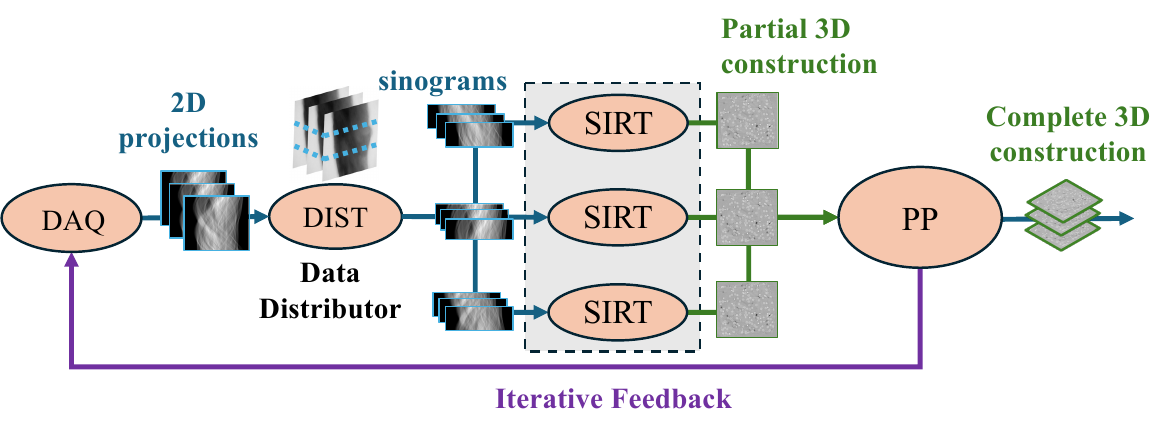}
    \caption{The Tomographic Reconstruction Workflow}
    \label{fig:aps-mini-apps}
\end{figure}

\subsection{Methodology}

\paragraph{\bf Objectives}

The evaluation assesses whether StreamGuard improves reliability in scientific streaming workflows without compromising performance. Our methodology is designed to isolate the impact of resilience mechanisms from application behavior and system noise, allowing us to evaluate both effectiveness and efficiency under controlled conditions. Specifically, we evaluate the following aspects:

\begin{enumerate}[topsep=0pt,itemsep=0pt,leftmargin=*]
    \item \textit{Resilience Capability}: whether StreamGuard mitigates failures and performance anomalies sufficiently to allow workflows to complete within real-time constraints;

    \item \textit{Performance Transparency}: whether StreamGuard introduces minimal overhead during failure-free execution;

    \item \textit{Robustness}: whether StreamGuard provides consistent protection across different scales, failure rates, and anomaly scenarios.
\end{enumerate}

Together, these objectives evaluate whether resilience can be achieved as a practical system property rather than a trade-off against performance.

\paragraph{\bf Metrics}

\begin{figure*}[t]
    \centering
    \subfloat[No failures]{
        \includegraphics[width=0.33\linewidth, trim={0 0 0 0}, clip]{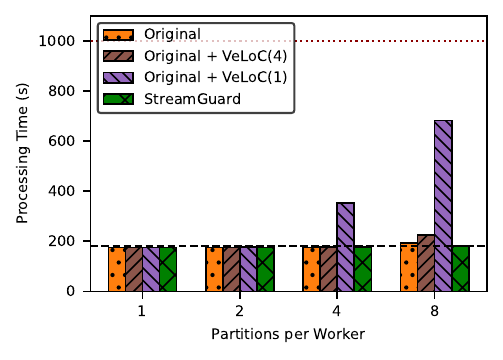}
        \label{fig:efficientcy-ckpt-no-failures}
    }
    \subfloat[MTTF = 150\,s]{
        \includegraphics[width=0.33\linewidth, trim={0 0 0 0}, clip]{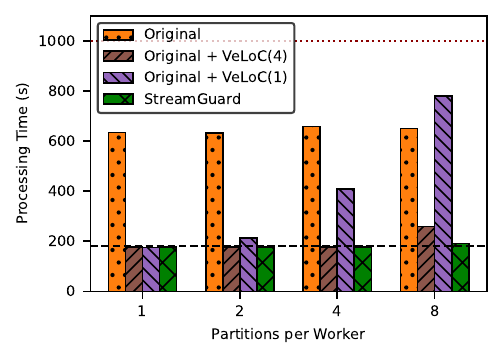}
        \label{fig:efficiency-ckpt-mttf-75}
    }
    \subfloat[MTTF = 40\,s]{
        \includegraphics[width=0.33\linewidth, trim={0 0 0 0}, clip]{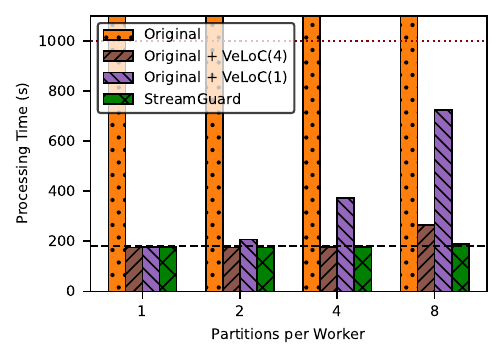}
        \label{fig:efficiency-ckpt-mttf-20}
    }
    \caption{Dynamic checkpointing efficiency. StreamGuard maintains processing time close to ideal execution (black dashed line) and consistently meets the real-time deadline (red dotted line) across varying computation loads and failure scenarios. In contrast, the original implementation fails to meet the deadline under failures, while static checkpointing depends on preconfigured intervals, resulting in inconsistent performance as failure rates and workloads change.}
    \label{fig:efficiency-ckpt}
\end{figure*}

\begin{figure*}[t]
    \centering
    \subfloat[4 Workers]{
        \includegraphics[width=0.33\linewidth, trim={0 0 0 0}, clip]{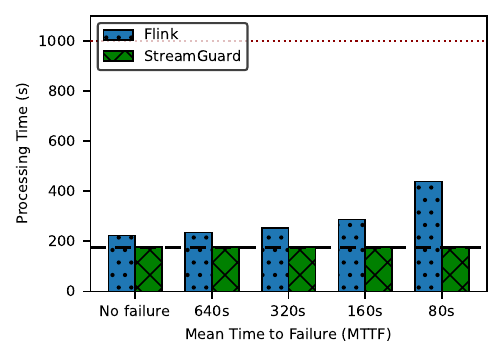}
        \label{fig:varying-mttf-worker-4}
    }
    \subfloat[8 Workers]{
        \includegraphics[width=0.33\linewidth, trim={0 0 0 0}, clip]{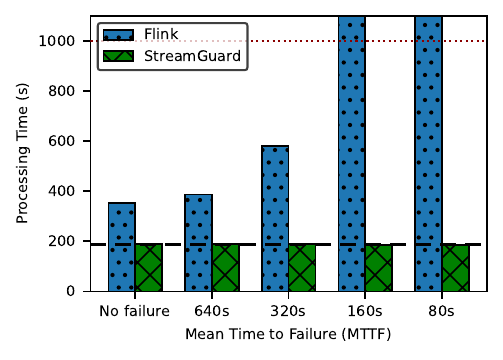}
        \label{fig:varying-mttf-worker-8}
    }
    \subfloat[16 Workers]{
        \includegraphics[width=0.33\linewidth, trim={0 0 0 0}, clip]{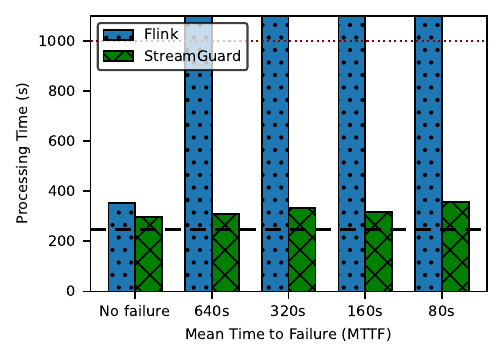}
        \label{fig:varying-mttf-worker-16}
    }
    \caption{
    Non-blocking asynchronous checkpoint efficiency. StreamGuard allows each consumer (reconstruction worker) to perform checkpoints independently and overlap checkpointing with computation, resulting in negligible overhead (with respect to ideal processing time, dashed black lines) and enabling real-time deadlines (red dotted lines) to be met across scales, even under very frequent failures. In contrast, Flink requires coordinated checkpointing through barrier propagation across workers, which increases synchronization overhead and causes deadline violations at high failure rates and large scales.
    }
    \label{fig:resilience-vs-mttf}
\end{figure*}

The above objectives are evaluated using metrics reflecting the operational requirements of scientific streaming: (1) \textit{Processing Time}, measured from the first producer emission to the last consumer output, capturing both the resilience and efficiency of the solution; (2) \textit{Deadline Violation}, whether processing time exceeds the maximum acceptable limit; and (3) \textit{Overhead}, the additional time relative to ideal failure-free execution, quantifying performance transparency.

\paragraph{\bf Workload}

We evaluate StreamGuard using a real-world scientific streaming workflow based on the Trace reconstruction engine~\cite{bicer2017trace,bicer2020tomographic}, which reconstructs cross-sectional images from X-ray projections collected at synchrotron radiation facilities. This workload is representative of scientific streaming applications due to its iterative computation, strict timing requirements, and sensitivity to failures and performance imbalance.

As shown in \autoref{fig:aps-mini-apps}, the workflow consists of four stages. The data acquisition component (\texttt{DAQ}) collects projection images of size $2048 \times 2048$ pixels from synchrotron radiation facilities. The distributor (\texttt{DIST}) normalizes collected data, converts projections into sinograms, and distributes them to parallel reconstruction workers (\texttt{SIRT}), which perform iterative partial 3D reconstruction. The post-processing stage (\texttt{PP}) aggregates partial results into complete reconstructed images.

Our evaluation focuses on the DIST–SIRT producer–consumer pair because it dominates per-iteration cost, and is where reconstruction state is formed, so failures and slowdowns there most directly affect end-to-end latency.
To reflect operational constraints, the pair's outputs must be produced within a fixed deadline. Under normal execution, reconstruction requires approximately 300--350 iterations over about 1500 records, resulting in an average processing time of roughly 200\,s. We therefore set the deadline to 1000\,s, providing sufficient recovery margin while preserving realistic real-time requirements.

\paragraph{\bf Baselines}
\label{sec:evaluation-method-baselines}

We compare StreamGuard against baselines covering both HPC-style and stream-based resilience mechanisms:

\begin{enumerate}[topsep=0pt,itemsep=0pt,leftmargin=*]
    \item \textit{Original Implementation}. The unmodified workflow without resilience mechanisms. This configuration represents the performance upper bound under failure-free execution but provides no protection against failures or anomalies.

    \item \textit{Original + VeloC ($X$)}. The original workflow augmented with VeloC \cite{VELOC-SuperCheck21} checkpointing every $X$ iterations; a lightweight, state-of-the-art HPC baseline that enables comparison of static against dynamic checkpoint policies.

    \item \textit{Flink}. The workflow is implemented using Apache Flink, a widely used stream processing engine with built-in fault tolerance for stateful operators. Flink maintains resilience by inserting checkpoint barriers into data streams; workers pause processing upon receiving barriers, snapshot their state, and propagate barriers downstream until a globally consistent checkpoint is established. Since Flink does not natively restart failed workers, we implement an external monitoring mechanism to detect failures and relaunch workers when necessary. We further tune checkpointing modes, data partitioning, and parallelism to obtain Flink’s best-performing configuration for fair comparison.

    \item \textit{StreamGuard}. Implementing our resilient mechanisms as described in \autoref{sec:implementation}. Although the resilience components are designed as independent modules, they are enabled together during evaluation whenever applicable. This allows us to assess their combined impact in realistic deployment scenarios, capturing both the overall resilience benefits and the cumulative overhead introduced by the complete solution.
\end{enumerate}

\paragraph{\bf System Configurations}
\label{sec:evaluation-method-system}

All experiments are conducted on the Polaris supercomputer at Argonne National Laboratory. To minimize external interference, the workflow is deployed on dedicated compute nodes without co-located applications. Each node contains a single AMD EPYC ``Milan'' processor with 64 physical cores and 512\,GB of memory, interconnected via a 200\,Gbps high-speed network. Checkpoints are stored on the Lustre parallel file system, enabling shared access during recovery.

Failures and anomalies are injected using independent modules that operate transparently to the workflow. For failure injection, a daemon periodically terminates active workers by sending \texttt{kill} signals, with inter-failure intervals drawn from an exponential distribution, which is commonly used to model real-world failure behavior \cite{dongarra2015fault}. The distribution is parameterized by a predefined mean time to failure (MTTF), allowing controlled evaluation under different failure intensities.

To emulate performance anomalies, each worker is assigned an artificial slowdown that adds additional computation before processing each record. The slowdown magnitude varies across workers and is sampled from an exponential distribution, commonly used to approximate the power-law behavior observed in performance variability and straggler effects \cite{dean2013tail}. The distribution is parameterized by a predefined mean at the start of each run, and the assigned slowdown remains fixed for the worker's lifetime.

\subsection{Experimental Results}
\label{sec:evaluation-results}

\begin{figure*}[t]
    \centering
    \subfloat[4 workers]{
        \includegraphics[width=0.33\linewidth, trim={0 0 0 0}, clip]{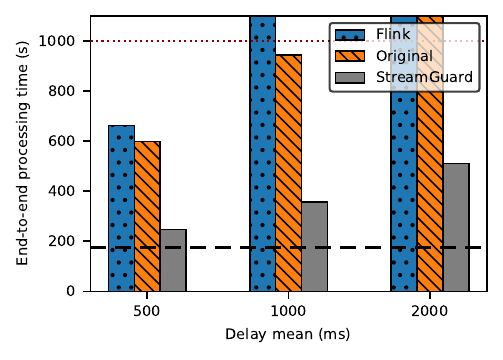}
        \label{fig:delay-sirt-4}
    }
    \subfloat[8 workers]{
        \includegraphics[width=0.33\linewidth, trim={0 0 0 0}, clip]{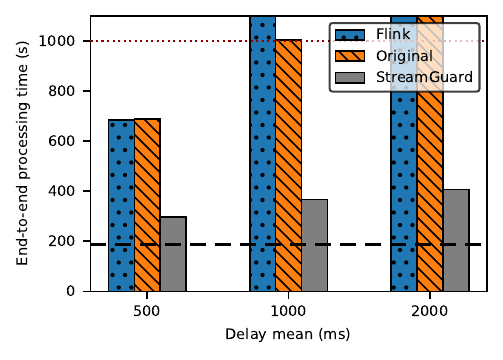}
        \label{fig:delay-sirt-8}
    }
    \subfloat[16 workers]{
        \includegraphics[width=0.33\linewidth, trim={0 0 0 0}, clip]{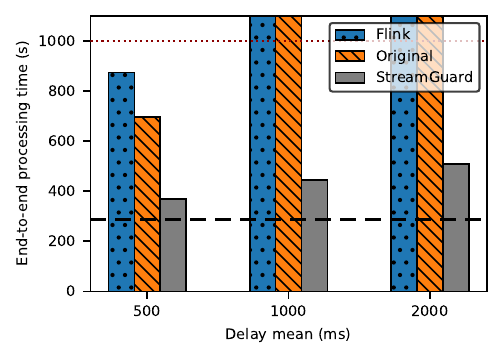}
        \label{fig:delay-sirt-16}
    }
    \caption{Dynamic load-balancing efficiency. Load rebalancing based on worker speed and per-partition progress significantly reduces processing lag caused by slow workers, enabling real-time constraints to be satisfied across scales and anomaly scenarios where existing solutions fail. The black dotted line indicates the processing time of the original implementation under failure-free (ideal) execution, while the red dotted line denotes the processing deadline.}
    \label{fig:resilience-load-balancing}
\end{figure*}

\begin{figure*}[t]
    \centering
    \subfloat[No delay]{
        \includegraphics[width=0.33\linewidth, trim={0 0 0 0}, clip]{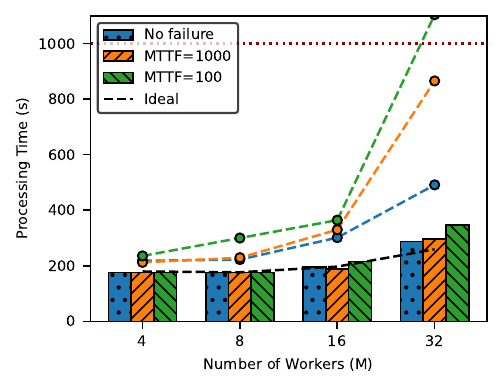}
        \label{fig:scalability-no-delay}
    }
    \subfloat[Mean delay = 500\,ms]{
        \includegraphics[width=0.33\linewidth, trim={0 0 0 0}, clip]{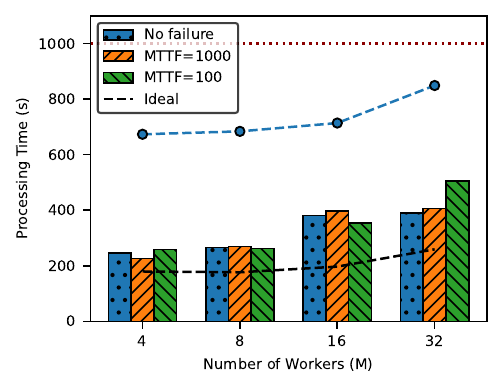}
        \label{fig:scalability-500ms-delay}
    }
    \subfloat[Mean delay = 1000\,ms]{
        \includegraphics[width=0.33\linewidth, trim={0 0 0 0}, clip]{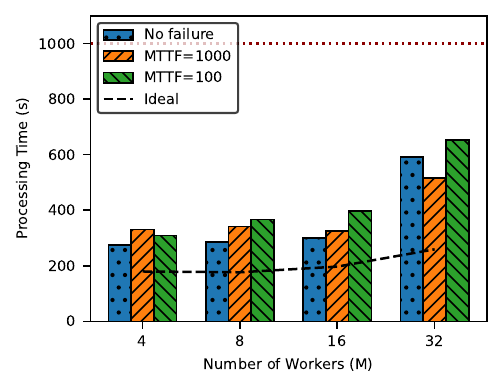}
        \label{fig:scalability-1000ms-delay}
    }
    \caption{
    Resilience robustness. StreamGuard maintains acceptable processing time and satisfies all real-time deadlines across failure rates, anomaly conditions, and scales. Bars show StreamGuard processing time; lines show Flink processing time. Colors indicate failure settings. Missing lines indicate Flink runs that exceed the deadline.
    }
    \label{fig:resilience-robustness}
\end{figure*}

\paragraph{\bf Checkpoint and Retry Efficiency}

First, we evaluate the effectiveness of StreamGuard's checkpoint and retry mechanism in protecting streaming data processing against unexpected failures and compare its performance with state-of-the-art solutions from both HPC and industrial stream-processing domains, represented by VeloC and Flink, respectively.

\autoref{fig:efficiency-ckpt} reports consumer processing time as we vary the number of data partitions mapped to each worker under three conditions: failure-free execution, light failures (per-worker MTTF = 150\,s), and heavy failures (MTTF = 40\,s). The key observation is that StreamGuard's asynchronous, non-blocking checkpoint mechanism introduces negligible overhead in failure-free runs. As shown in \autoref{fig:efficientcy-ckpt-no-failures}, the overhead remains below 1\%, and the completion time is nearly identical to execution without checkpointing (green bars). This demonstrates that resilience can be achieved without sacrificing performance when failures do not occur.

In contrast, the checkpoint-free configuration in the original implementation (orange bars) is highly vulnerable to failures. Each failure forces a worker to restart from the beginning, discarding previously completed computation and significantly increasing total processing time. Even under a light failure rate (MTTF = 150\,s), recomputation overhead becomes substantial (\autoref{fig:efficiency-ckpt-mttf-75}). Under heavy failures (MTTF = 40\,s, \autoref{fig:efficiency-ckpt-mttf-20}), the accumulated recomputation cost becomes prohibitive, preventing the workflow from completing within the deadline window (1000\,s).

Static checkpointing cannot fully resolve this problem. Frequent checkpointing (\texttt{Original+VeloC(1)}, purple bars) sustains progress under heavy failures but incurs excessive overhead when failures are infrequent, exceeding 3$\times$ at eight partitions per worker due to I/O contention from concurrent checkpoints. Reducing frequency (\texttt{Original+VeloC(4)}, brown bars) lowers failure-free overhead but becomes ineffective at high failure rates. A fixed interval cannot adapt to changing failure rates and load. In contrast, StreamGuard adapts checkpoint frequency to observed cost and failure behavior, maintaining low overhead and resilience across all scenarios.

\autoref{fig:resilience-vs-mttf} compares StreamGuard with Flink, which employs synchronized, barrier-based checkpointing. This mechanism places checkpoint cost directly on the processing critical path because workers must pause while snapshots are taken. Moreover, checkpoint progress depends on barrier propagation across all workers; under failures or performance imbalance, slower workers delay barrier completion, causing congestion and backpressure. As shown in \autoref{fig:varying-mttf-worker-4}, Flink execution is already 25\% slower than the original implementation under failure-free conditions, and processing time increases further as failure frequency grows. The degradation is magnified at scale: at 16 workers with per-worker MTTF of 640\,s, the SIRT stage experiences disruptions approximately every 40\,s, preventing checkpoint barriers from reliably traversing the workflow and causing deadline violations.

In contrast, StreamGuard performs checkpoints in a fully non-blocking and asynchronous manner, removing checkpoint overhead from the processing critical path. In addition, data are streamed through a dedicated queueing service (Mofka), allowing workers to fetch data on demand and decoupling resilience behavior from flow control and data movement. As a result, checkpoint progress remains unaffected by transient slowdowns or failures in other workers. Across all evaluated scales and failure settings in \autoref{fig:resilience-vs-mttf}, StreamGuard consistently meets real-time deadlines. Even under the most extreme case with 16 workers and MTTF = 80\,s, which is equivalent to one failure every five seconds within the \texttt{SIRT} stage, the processing time remains within 20\% of the ideal execution.

\vspace{2mm}
\noindent
\colorbox{blue!10}{
\parbox{0.96\linewidth}{
\underline{\textbf{Takeaway}:} \emph{
Dynamic, non-blocking, asynchronous checkpoint and retry minimizes failure impact while preserving near-ideal performance, enabling robust real-time processing, even under extreme conditions, where static or synchronized checkpointing fails.
}
}}

\paragraph{\bf Dynamic Load Balancing}

Next, we evaluate the effectiveness of StreamGuard's dynamic load-balancing algorithm in mitigating the impact of slowdown anomalies. To emulate heterogeneous worker performance, we deploy the workflow with injected slowdown using the anomaly model described in \autoref{sec:evaluation-method-system}. \autoref{fig:resilience-load-balancing} reports the processing time of the workflow under StreamGuard compared with Flink and the original implementation across three slowdown configurations: \textit{light} (average slowdown of 500\,ms), \textit{medium} (1000\,ms), and \textit{heavy} (2000\,ms), and across three deployment scales of 4, 8, and 16 workers.

Both Flink and the original implementation rely on static partition-to-worker mappings. As a result, end-to-end processing time is dictated by the slowest worker. When slowdown anomalies are present, this leads to substantial performance degradation, with processing time increasing by at least 3$\times$. The impact becomes more pronounced as the number of workers increases, since a larger deployment increases the likelihood of encountering slower workers that delay overall progress. Under medium and heavy slowdown conditions (1000\,ms and 2000\,ms), the accumulated delay prevents both approaches from completing processing before the deadline, even at small scale with only four workers.

In contrast, StreamGuard continuously monitors partition progress and estimates effective worker speed during execution. Lagging partitions are proactively migrated to faster workers, allowing the system to rebalance workload dynamically as performance imbalance emerges. This dynamic rebalancing progressively equalizes progress across partitions and prevents slow workers from dominating overall execution time. As a result, with four workers, processing time increases by at most 3$\times$ while still meeting the real-time deadline, whereas both Flink and the original implementation fail to complete on time. Furthermore, as the number of workers increases, processing time remains stable because additional faster workers can compensate for slower ones. This allows StreamGuard to maintain nearly constant overhead across scales and consistently satisfy real-time constraints.

\vspace{2mm}
\noindent
\colorbox{blue!10}{
\parbox{0.96\linewidth}{
\underline{\textbf{Takeaway}:} \emph{Dynamic load balancing effectively mitigates slowdown anomalies by carefully reassigning slow partitions to fast workers, enabling robust real-time processing across scales and heterogeneous execution conditions.}
}}

\paragraph{\bf Robustness}

Finally, we evaluate the robustness of StreamGuard by executing the workflow under combined failure and slowdown anomaly conditions at multiple scales, and comparing the results against both Flink and the ideal execution. The results are summarized in \autoref{fig:resilience-robustness}. We consider three anomaly settings: no slowdown (\autoref{fig:scalability-no-delay}), an average slowdown of 500\,ms (\autoref{fig:scalability-500ms-delay}), and an average slowdown of 1000\,ms (\autoref{fig:scalability-1000ms-delay}). Each anomaly setting is evaluated under three failure configurations: no failures (blue), per-worker MTTF of 1000\,s (orange), and MTTF of 100\,s (green), while scaling the deployment from four to 32 workers. Bars represent the processing time of StreamGuard, while lines represent Flink processing time.

Consistent with observations from previous experiments, Flink’s synchronized checkpointing mechanism makes it increasingly sensitive to high failure rates, particularly at larger scales where the probability of failure increases with the number of workers. As scale and failure frequency increase, checkpoint coordination overhead grows, resulting in longer processing times. The presence of slowdown anomalies further amplifies this effect by extending processing duration, which in turn exposes Flink to additional failures within the same execution window. This compounding effect leads to rapidly increasing overhead, causing Flink to miss the processing deadline in most configurations once the average slowdown reaches 500\,ms.

In contrast, StreamGuard maintains stable performance across all tested configurations. Failures and slowdown anomalies are handled by separate resilience mechanisms with minimal coupling, preventing their negative effects from amplifying each other. Dynamic checkpointing limits recovery overhead under frequent failures, while load balancing mitigates processing lag caused by slow workers. As a result, processing time remains close to the ideal execution even as scale, failure rate, and anomaly severity increase, allowing the workflow to consistently meet real-time deadlines across all evaluated settings.

\vspace{2mm}
\noindent
\colorbox{blue!10}{
\parbox{0.96\linewidth}{
\underline{\textbf{Takeaway}:} \emph{By handling failures and anomalies independently with minimal cross-dependency, StreamGuard prevents their effects from compounding, enabling robust real-time execution across a range of failure rates, anomaly conditions, and deployment scales.}
}}

\section{Conclusions and Future Work}

We presented StreamGuard, a resilience architecture for real-time scientific streaming workflows. By treating the producer–consumer pair as the unit of resilience and integrating dynamic non-blocking checkpointing with progress-aware load balancing as decoupled, composable modules, StreamGuard avoids global coordination while maintaining real-time deadlines under frequent failures, persistent slowdown anomalies, and increasing scale. Experiments on a real-world tomographic reconstruction workflow show that resilience can be achieved as a first-class system property without sacrificing scalability or performance transparency.

Several promising directions extend this work. First, StreamGuard assumes approximately uniform per-partition computational cost, which may not always hold in practice. Extending the load balancer's capacity model with per-partition cost weights would broaden applicability to skewed workloads. Second, StreamGuard is evaluated with sufficient allocated resources for the experimental workload. Yet in practice, load spikes or miscalculations can cause transient under- or over-provisioning. Although our prior work has studied resource-workload mismatch \cite{nguyen2025resilient}, its setting does not target real-time streaming; studying such scenarios in a streaming context would help develop more robust and practical solutions. Finally, while the mechanisms presented here are stage-local by design, evaluating compound effects across multiple producer-consumer stages, including cross-stage I/O interference and backpressure, remains an important direction for future work.

\begin{acks}
This work was supported in part by the U.S.\ Department of Energy under Contract DE-AC02-06CH11357, as well as the National Science Foundation (NSF), Office of Advanced Cyberinfrastructure, under grants CSSI-2411386 and CSSI-2514056.
\end{acks}

\bibliographystyle{ACM-Reference-Format}
\bibliography{references}

\end{document}